\documentclass[12pt]{article}

\usepackage{arxiv}

\usepackage[utf8]{inputenc} 
\usepackage[T1]{fontenc}    
\usepackage{hyperref}       
\usepackage{url}            
\usepackage{booktabs}       
\usepackage{amsfonts}       
\usepackage{nicefrac}       
\usepackage{microtype}      
\usepackage{lipsum}
\usepackage{graphicx}
\usepackage{amssymb,amsfonts,amsmath,amsthm}
\usepackage{algorithm}
\usepackage{algpseudocode}
\usepackage{tikz}
\usepackage{caption}
\usepackage{subcaption}
\usepackage{textcomp}
\usepackage{xcolor}
\usepackage[english]{babel}

\usepackage{authblk}

\title{Intrinsic Hierarchical Clustering Behavior Recovers Higher Dimensional Shape Information}

\author{Paul Samuel Ignacio}

\affil{University of the Philippines Baguio, Baguio City, Philippines 2600\\ppignacio@up.edu.ph}

\begin{document}
\date{}
\maketitle
\begin{abstract}
We show that specific higher dimensional shape information of point cloud data can be recovered by observing lower dimensional hierarchical clustering dynamics. We generate multiple point samples from point clouds and perform hierarchical clustering within each sample to produce dendrograms. From these dendrograms, we take cluster evolution and merging data that capture clustering behavior to construct simplified diagrams that record the lifetime of clusters akin to what zero dimensional persistence diagrams do in topological data analysis. We compare differences between these diagrams using the bottleneck metric, and examine the resulting distribution. Finally, we show that statistical features drawn from these bottleneck distance distributions detect artefacts of, and can be tapped to recover higher dimensional shape characteristics.
\end{abstract}

\section{Introduction}
\label{intro}
Understanding the overall shape of point clouds is a central goal in topological data analysis. The basic task is to recover invariant signatures that characterize either the overall structure, or local properties (manifold learning) of the space where the point cloud is understood to be sampled from. The reward is that these signatures do not only afford us with a holistic understanding of the data's composition, but also meaningful information about the underlying data that would otherwise be missed by other methods. For example, Perea and Harer \cite{perea} used sliding window embeddings to examine time series gene expression data as point clouds and were able to detect periodicity in the presence of damping missed by state-of-the-art methods. The same strategy was employed in \cite{ignacio} where topology-based features are used as proxy for expert-dependent features to improve diagnosis of Atrial Fibrillation.

This idea of interrogating data via its underlying structure is not unique to topological data analysis. Traditionally, insights from clustering algorithms are used to describe the extent to which local or global regions of data tend to group together, or in many other settings to describe the connectivity of an associated graph. This grouping behavior among points, driven by pairwise measures of dissimilarity, provides a lowest dimension characterization of the overall shape of point cloud data. 

\begin{figure}
\centering
  \includegraphics[width=0.7\textwidth, height=0.35\textwidth]{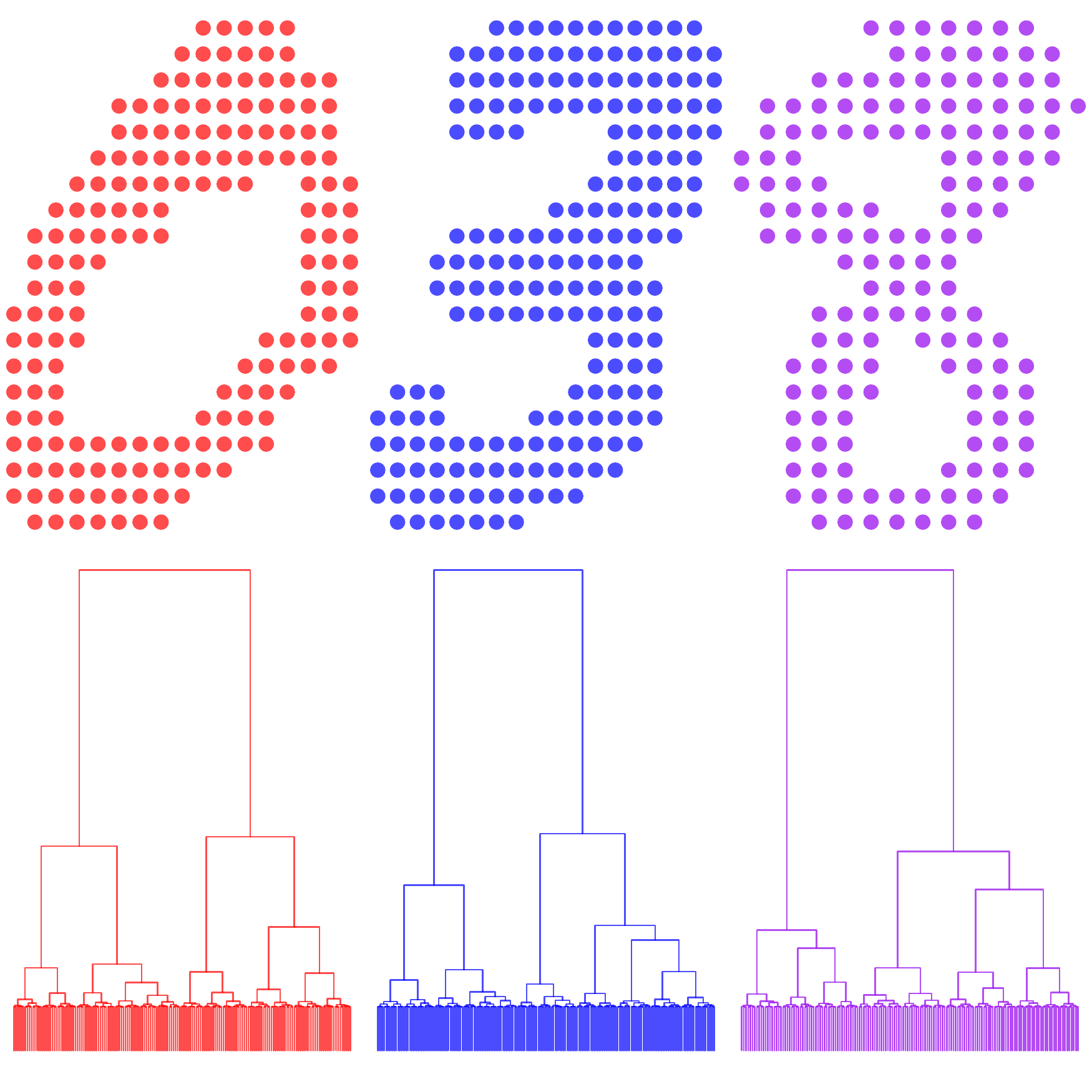} 
\caption{Point clouds sampled from the digits 0, 3, and 8 with corresponding dendrograms from hierarchical clustering.}
\label{pointclouds}       
\end{figure}

Clustering information, however, is very seldom enough to characterize the shape of point clouds as multiple point clouds may elicit the same overall clustering characteristics while having drastically different overall shapes. A simple example would be if one considers sufficiently sampled points from the digits 0, 3 and 8 considered as point cloud objects. Clustering the points independently in each sample reveals that all three have a single significant cluster (see Figure \ref{pointclouds}). While density-based clustering may be able to separate these three when all are embedded together in a common space, clustering independently using only intrinsic information within each sample would give very similar characterization for all three point clouds. To circumvent this limitation, the need to appeal to higher dimensional characterization, such as presence of loops or cycles, arises.

Access to higher dimensional characterization, however, comes at a cost as most methods suffer from combinatorial explosion as the size and complexity of data scale up. Unfortunately, the speed at which technology in this area is progressing is simply outpaced by the speed at which data can be collected. This setting, thus, challenges us to explore other ways of exploiting currently available methods and technology in ways that aid the state-of-the-art. For this, we turn our attention to clustering methods. 

Consider again the dendrograms below the digits in Figure \ref{pointclouds}. Observe that while the dominant characteristics of all three dendrograms are similar, the merging dynamics among the clusters in the three digits are actually different. It is this subtle differences in clustering behavior that we want to interrogate. In more ways than one, this approach is inspired by the work of Bubenik et al. in \cite{bubenik}, where they showed that specific portions of the summary produced by persistent homology, widely believed and accepted to be artefacts of noise, can actually be used to recover curvature in the underlying space. In particular, we follow their idea of utilizing unprocessed information from available output summaries to extract supplemental information, and contextualize this approach in the recovery of expensive higher dimensional information. In this work, the main question that we ask is ``\emph{Does intrinsic clustering behavior contain signal about higher dimensional shape characteristics?}"

We provide evidence to positively answer this question, and show that it is indeed possible to recover specific higher dimensional shape information of point clouds by observing clustering behavior across multiple point samples. This approach leverages intrinsic hierarchical clustering dynamics within point clouds to detect and retrieve artefacts of higher dimensional characteristics that clustering algorithms, by themselves, are agnostic to. Exacting empirical proof that this method works affords access to the inverse problem of using lower dimensional information, which are considerably easier, cheaper and faster to compute, to either be used as basis for classifying point clouds with varying higher dimensional characteristics, or to explicitly recover these higher dimensional shape information.

\section{Pipeline and Methods}
\label{sec:1}
\begin{figure}[h]
\centering
 \includegraphics[width=\textwidth,height = 0.35\textwidth]{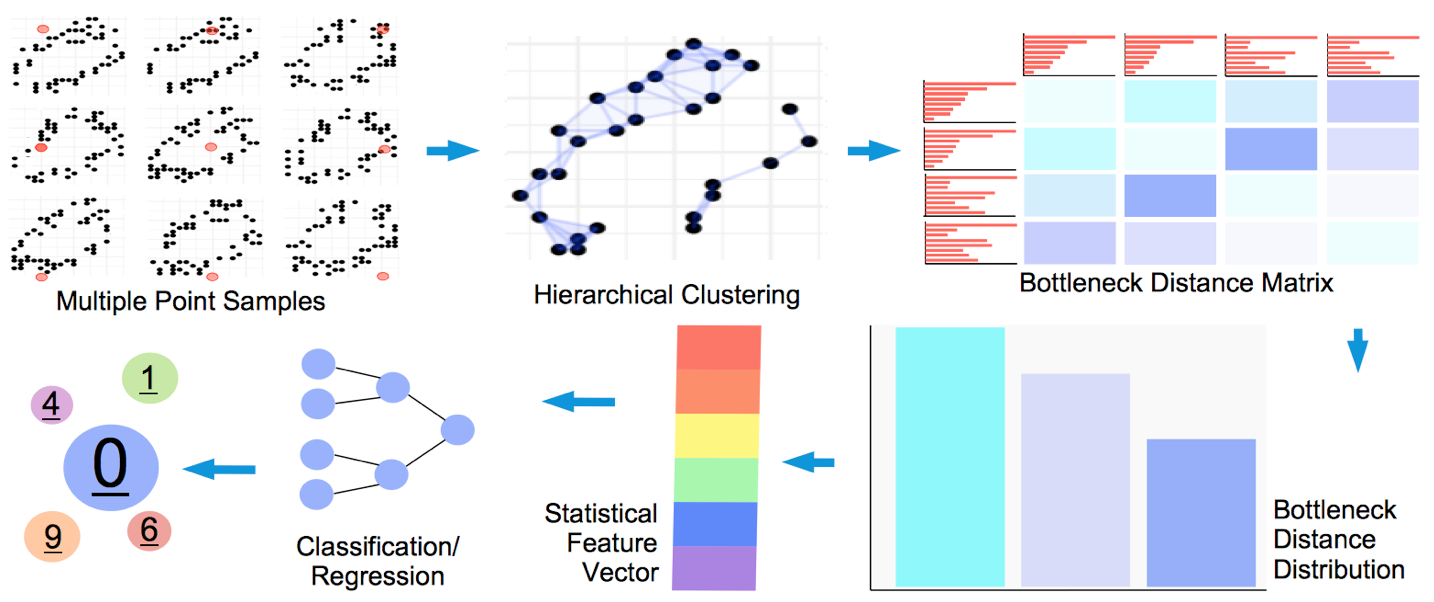}
\caption{Pipeline for recovering higher dimensional shape information using observed clustering behavior.}
\label{Pipeline}       
\end{figure}

The first step in our approach is to generate multiple collections of points sampled from a given point cloud. In this work, we consider as point clouds images of handwritten digits from the popular MNIST data set \cite{lecun}. Each collection of sampled points is independently clustered using hierarchical methods, and captured clustering behavior is extracted from their resulting dendrograms. In particular, cluster evolution and lifetime data are used to construct diagrams akin to persistence diagrams in topological data analysis. These diagrams are then compared using the bottleneck metric to obtain an overall comparison of the captured clustering behavior across point samples. Distributions of these comparisons are examined, and summaries are extracted and tested for classification and prediction tasks. The entire pipeline is shown in Figure \ref{Pipeline}.  

\subsection{Point Cloud Sampling}
\label{sec:sampling}
\begin{figure}[h]
\begin{subfigure}{0.3\textwidth}
\includegraphics[width=\textwidth]{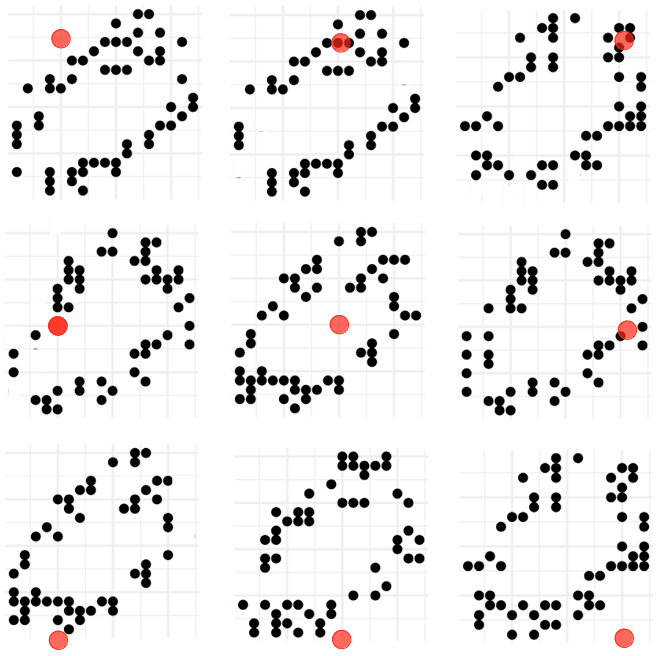}
\caption{}
\label{sample1}
\end{subfigure}
\begin{subfigure}{0.75\textwidth}
\centering
\noindent
\begin{tikzpicture}
\node [anchor=west,style={rotate=90}] (res) at (-0.1,0.25) {\large Resolution};
\begin{scope}
    \node[anchor=south west,inner sep=0] (image) at (0,0) {  \includegraphics[trim= 0 90 1 1,clip,width = 0.8\textwidth, height = 0.4\textwidth]{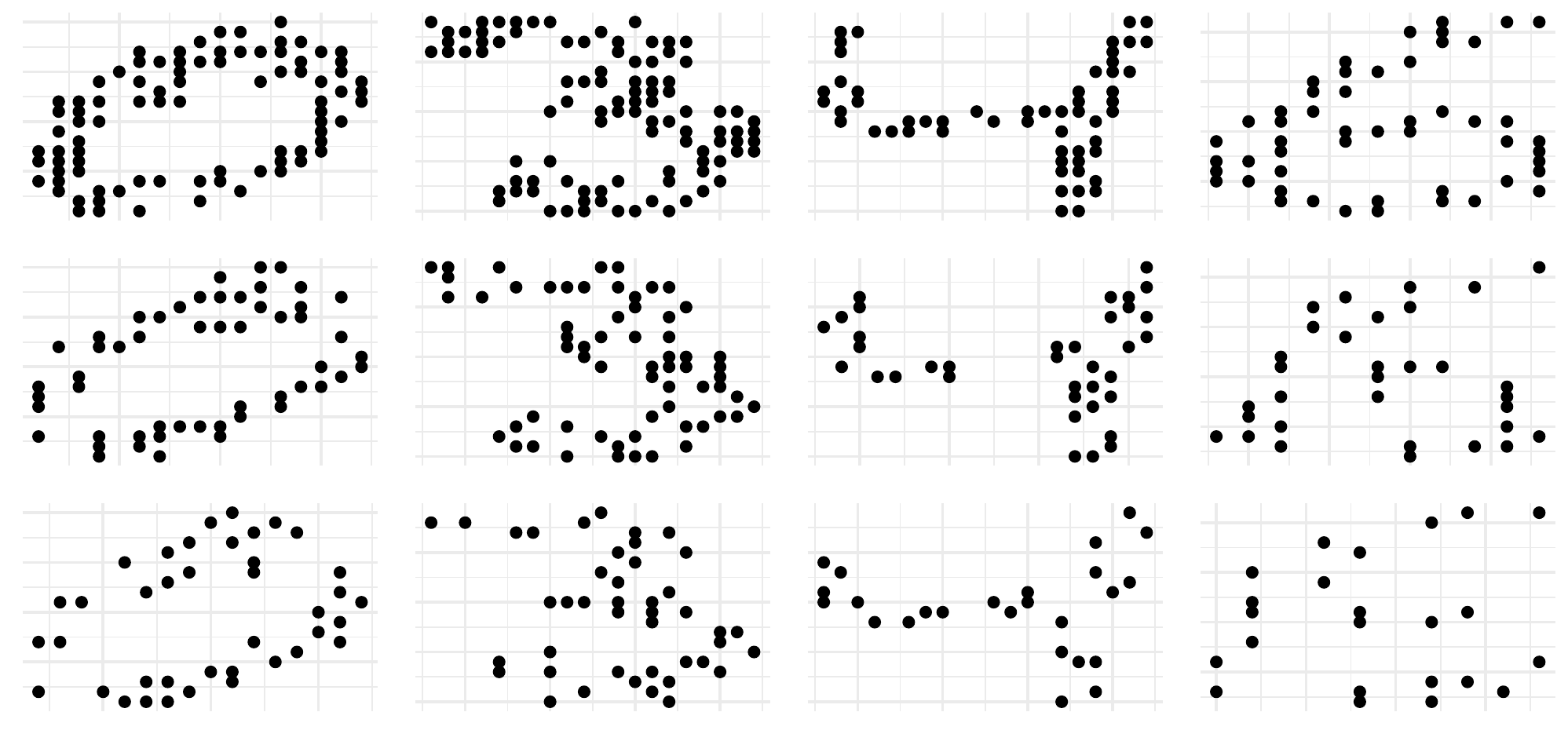}};
    \begin{scope}[x={(image.south east)},y={(image.north west)}]
        \draw [-stealth, line width=2pt, cyan] (res) -- ++(0.0,0.6);
    \end{scope}
\end{scope}
\end{tikzpicture}%
\caption{}
\label{sample2}
\end{subfigure}
\caption{(a) Points are sampled based on their distribution relative to landmark points shown in red. (b) Samples are also taken in varying levels of resolution.}
\label{sampling}       
\end{figure}

As the main idea of our approach is to leverage intrinsic characteristics observed from the clustering of sampled sub-collections of points, it is important that each sample is a good representation of the original point cloud. A caveat, however, is that observation is perspective-dependent. Thus, to minimize the overall effect of bias introduced from specific observations of clustering behavior, we capture such from multiple perspectives. We do this by extracting point samples that reflect the distribution of points relative to multiple pre-selected landmark locations in the ambient space where the point cloud lives (see Figure \ref{sampling}a). We use the same grid-based landmark points used by Garin and Tauzin \cite{garin} for their work on the same data set.

We also observe clustering behavior in multiple resolutions. Using the distribution of points relative to a pre-selected landmark, we consider several sampling resolutions in which the number of points selected per bin in the distribution histogram is $1/k$th of the bin size for $k = 2, 3, 4, 5, 6$, e.g. a sampling resolution for $k = 2$ selects $1/2$ of the total points in every histogram bin. This will allow us to capture clustering behavior in varying levels of granularity (see Figure \ref{sampling}b). We create $n=30$ instances for each of the 45 pairs of sampling resolution and landmark reference. We refer to each combination pair of resolution and landmark reference as a sampling setting.

\subsection{Clustering Diagrams and Bottleneck-based Summaries}
\label{sec:bottleneck}
\begin{figure}[h]
\centering
  \includegraphics[height=0.35\textwidth,width=0.8\textwidth]{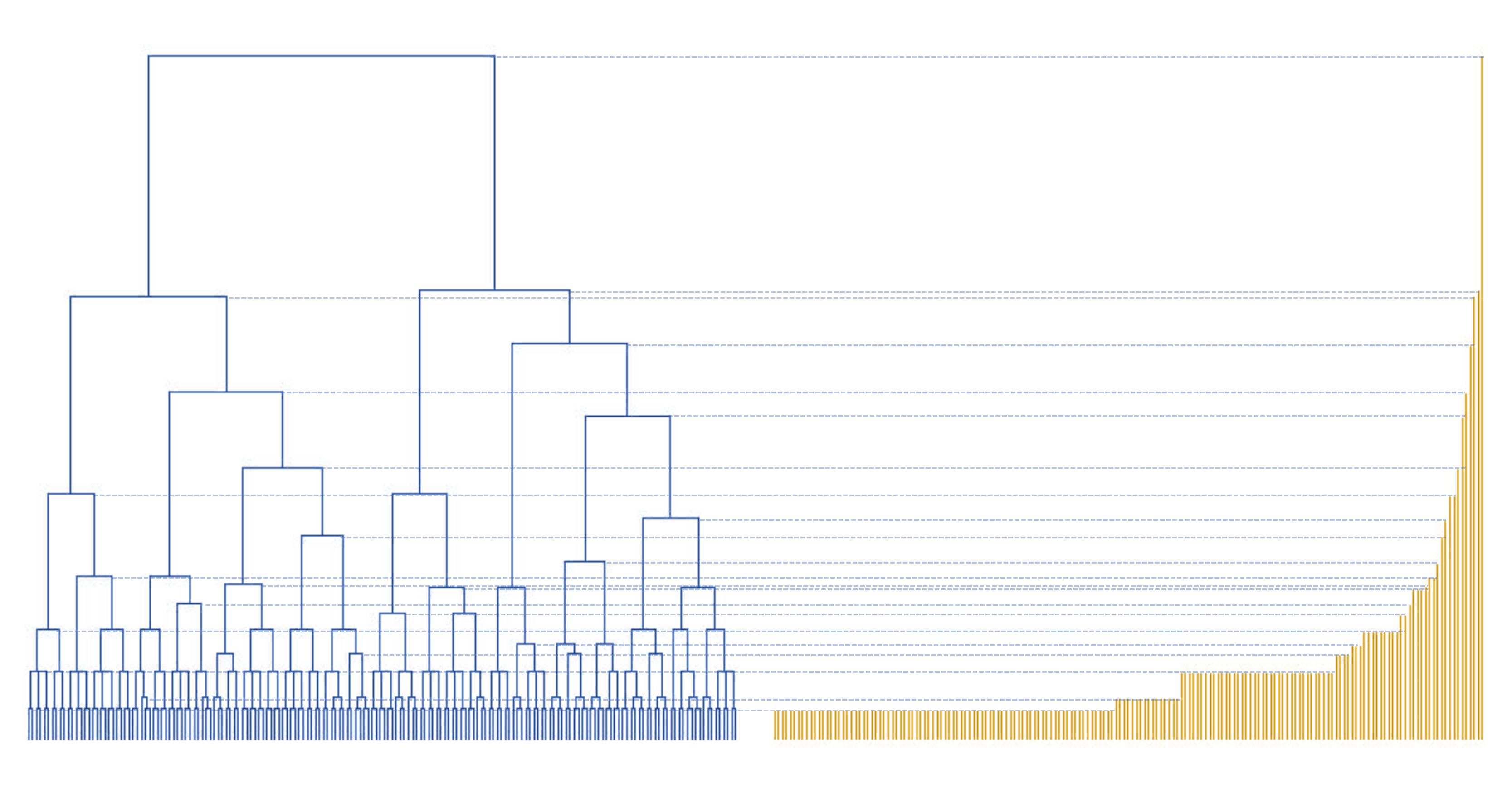} 
\caption{Clustering diagrams are constructed by extracting the merge heights of clusters in a dendrogram.}
\label{dendrogram}       
\end{figure}

For each sub-collection of sampled points, we use four hierarchical clustering algorithms to produce four distinct \emph{dendrograms}. A dendrogram is a tree-like multi-scale distance-parametrized record of the evolution of the clusters within a collection of points. Each leaf in the dendrogram represents a cluster and the merging of leaves the merging of the clusters they represent. The distance value at which clusters merge, and hence the clustering dynamics observed, is determined by the algorithm used. By employing several algorithms, we are able to record a nuanced observation of the clustering dynamics and evolution. It is this recorded history of cluster merging that we specifically employ hierarchical clustering algorithms. We would like to examine if features borne out of an examination of this cluster merging history contain recoverable artefacts of higher dimensional characteristics. The four hierarchical clustering algorithms we use are based on Single, Average, Complete, and Ward's linkage. For a quick introduction on these linkage methods, we refer the interested reader to \cite{jain}. To extract the clustering behavior, we take the merge heights $\Lambda := \{d_i\}_{i\in I}$ in the dendrogram and use it to construct a multi-set of points in the plane $\{(0,d)|d\in\Lambda\}$. To ease visualization, we illustrate this multi-set of points as a collection of bars with length corresponding to merge heights in the dendrogram  (see Figure \ref{dendrogram}). This multi-set of points, called a \emph{clustering diagram} \cite{leyda}, filters out unnecessary choices needed to properly illustrate a dendrogram while keeping important information that we want to explore. Clustering diagrams record the lifetime of clusters akin to what persistence diagrams in dimension zero do in topological data analysis. In fact, the clustering diagram constructed from the single linkage dendrogram recovers the dimension zero persistence barcode produced by the Vietoris-Rips filtration. 

To obtain a measure on the differences among the clustering behaviors observed across the sub-collections of points, we compute the pairwise differences across all clustering diagrams via the \emph{bottleneck distance} implemented in \textsc{Lum\'awig} \cite{lumawig}. Given two multi-set of points $X$ and $Y$ in the plane, the bottleneck distance between them is defined as
$$
d_B(X,Y) = \inf_{\phi} \sup_{x\in X} ||x-\phi(x)||_{\infty}
$$
where the infimum is taken over all bijections $\phi:X\sqcup \Delta \to Y\sqcup \Delta$ and $\Delta$ is the diagonal. In general terms, the bottleneck distance measures the cost to transform one multi-set to another.

We compute the distribution of pairwise bottleneck distances among the 30 clustering diagrams produced in each sampling setting. This distribution provides an insight on how observed clustering behaviors vary across the sampled representations of the original point cloud. From this distribution, we obtain the following statistical summaries:
\begin{enumerate}
    \item Minimum and maximum bottleneck distance values;
    \item Mean, standard deviation, skewness, and kurtosis of all bottleneck distances;
    \item Size of the largest bin in the histogram obtained from the distribution.
\end{enumerate}

\subsection{Higher Dimensional Shape Information via Persistent Homology}

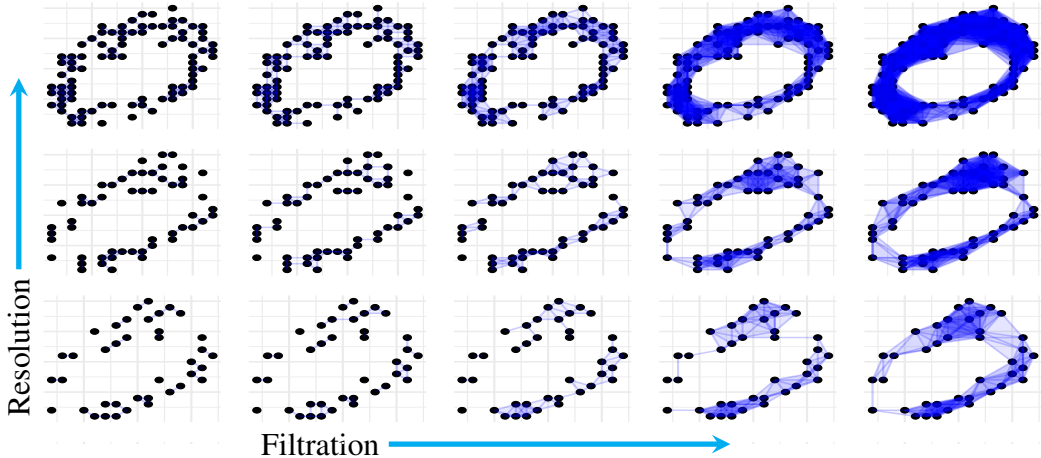
\begin{figure}
\centering
\noindent
\begin{tikzpicture}
\node [anchor=west,style={rotate=90}] (res) at (-0.1,0.25) {\large Resolution};
\node [anchor=west] (water) at (3,0) {\large Filtration};
\begin{scope}
    \node[anchor=south west,inner sep=0] (image) at (0,0) {  \includegraphics[trim= 0 195 1 1,clip,width = 0.8\textwidth, height = 0.35\textwidth]{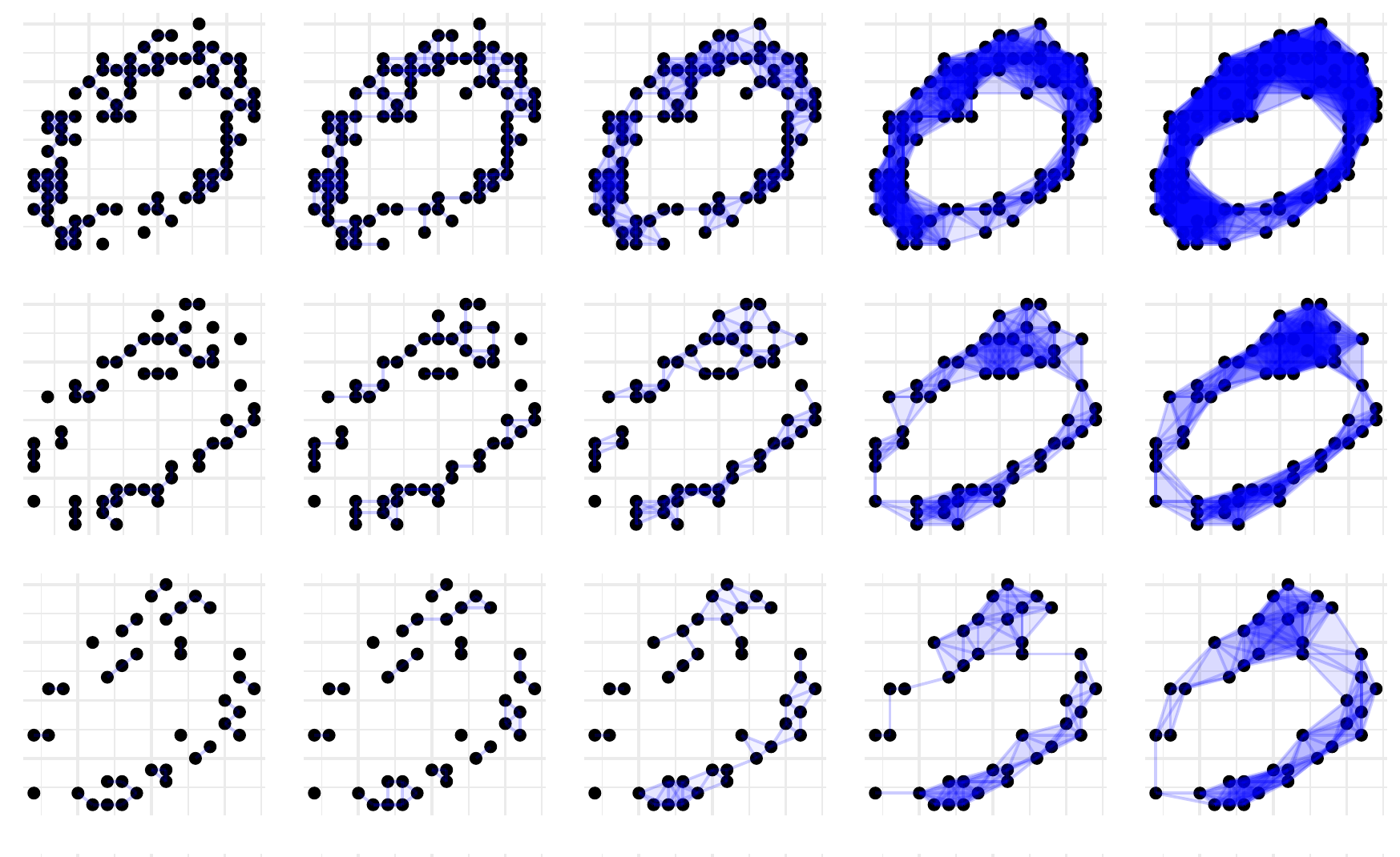}};
    \begin{scope}[x={(image.south east)},y={(image.north west)}]
        \draw [-stealth, line width=2pt, cyan] (res) -- ++(0.0,0.6);
        \draw [-stealth, line width=2pt, cyan] (water) -- ++(0.4,0.0);
    \end{scope}
\end{scope}
\end{tikzpicture}%
\caption{A Vietoris-Rips filtration is induced over each point cloud in every sampling setting.}
\label{filtration}       
\end{figure}

We generate higher dimensional characteristics of point clouds using persistent homology. For a quick introduction to persistent homology, we refer the interested reader to \cite{roadmap}. Succinctly, we induce a distance-parameterized sequence of simplicial complexes built over each point cloud sample to obtain mathematically computable topological signatures that persist across multiple scales (see Figure \ref{filtration}). In this approach, persistence over a large range of scales is regarded as a measure of significance for the corresponding topological characteristic. 

We extract persistent homological signatures pertaining to cycles (i.e. two-dimensional holes) in the point clouds. In particular, we take the number of detected persistent cycles, the average persistence, and the maximum persistence for each of the 30 sub-collections of points in every sampling setting, and take the mean for each feature. In what follows, we examine whether our clustering-based features can recover both the information provided by persistence-based higher dimensional features and their explicit values.

\section{Recovering higher dimensional shape information}
Our goal is to provide empirical proof that statistical summaries obtained from intrinsic clustering behavior contain information that can be used to successfully classify point clouds with varying higher dimensional characteristics, and explicitly predict values for such. We employ a \emph{random forest classifier} for the first task, and a \emph{random forest regressor} for the second task. Succinctly, a \emph{random forest} is an ensemble of \emph{decision trees} that uses the collective output of all trees in coming up with either a classification or a prediction. We train a 1000-tree random forest using an initial pool of 1,395 features that include clustering-based features and persistence-based higher dimensional features. For distinction, we refer to features coming from the first group as \emph{dimension 0} features and the second as \emph{dimension 1} features.

\subsection{Classification of Digit Images in MNIST}
\label{mnist}
 After an initial training, we rank all features from our initial pool by importance, and extract the top 200 features of which 150 are dimension 0 and 50 are dimension 1. We construct 3 training feature sets from these. The first set consists of only dimension 0 features, the second set consists of only dimension 1 features, and the third set consists of all 200 dimensions 0 and 1 features. To obtain comparable results from training across these 3 feature sets, no other fine tuning is done for the random forest training. We perform a 30-fold stratified cross validation on the full MNIST training data set of 60,000 digit images and report the performance in Table \ref{classifiercross}. We discuss first the performance of the random forest trained using the first feature set of only dimension 0 features.
 
 \begin{table}[h]
\centering
\caption{$F_1$ scores over a 30-fold stratified cross validation of the Random Forest trained on dimension 0 features to classify digits.}
\label{classifiercross}       
\begin{tabular}{cccccccccccc}
\hline\noalign{\smallskip}
Digit & 0 & 1 & 2 & 3 & 4 & 5 & 6 & 7 & 8 & 9 & Overall   \\
\noalign{\smallskip}\hline\noalign{\smallskip}
Mean & 0.777 & 0.927 & 0.649 & 0.653 & 0.644 & 0.629  & 0.791 & 0.741 & 0.699 & 0.705 & 0.722 \\
SD & 0.020 & 0.012 & 0.034 & 0.020 & 0.030 & 0.029 & 0.021 & 0.025 & 0.028& 0.030&0.011 \\
\noalign{\smallskip}\hline
\end{tabular}
\end{table}
 
Clusters are often interpreted as the components of point cloud data. Using linkage-based clustering algorithms, these clusters are further understood as connected components in the graph induced by the linking criteria, and the hierarchical nature of the algorithm suggests that the significant connected components of the underlying point cloud are captured by the long branches in the corresponding dendrogram. It is worth pointing out that despite the fact that all the digits represented in the MNIST data set have the same number of significant connected components (i.e. they all have a single connected component), the clustering behavior observed still provides a relatively good basis for classifying the digits as evidenced by the obtained $F_1$ scores. We take note in particular the relatively high scores for the digits $0, 6, 8,$ and $9$, which all have a higher dimensional topological characteristic of possessing  cycles\footnote{Although it can be argued that the digit ``4'' may also be considered in this class, handwriting styles introduce variations of this digit that do not contain the topological hole.}.

To put this performance in perspective, we investigate how much new information is contributed by dimension 1 features, and is unseen by our clustering-based dimension 0 features. We train the random forest using the two other feature sets and plot their performances against the performance when training with only dimension 0 features. The idea is that, as dimension 1 features capture higher dimensional characteristics present in several digit classes, they themselves could be used as basis for classification, and since clustering algorithms are agnostic to these higher dimensional characteristics that in theory are good discriminants for the digits, their inclusion should produce a significant increase across all class scores.

\begin{figure}[h]
\centering
  \includegraphics[height = 0.75\textwidth,width=0.8\textwidth]{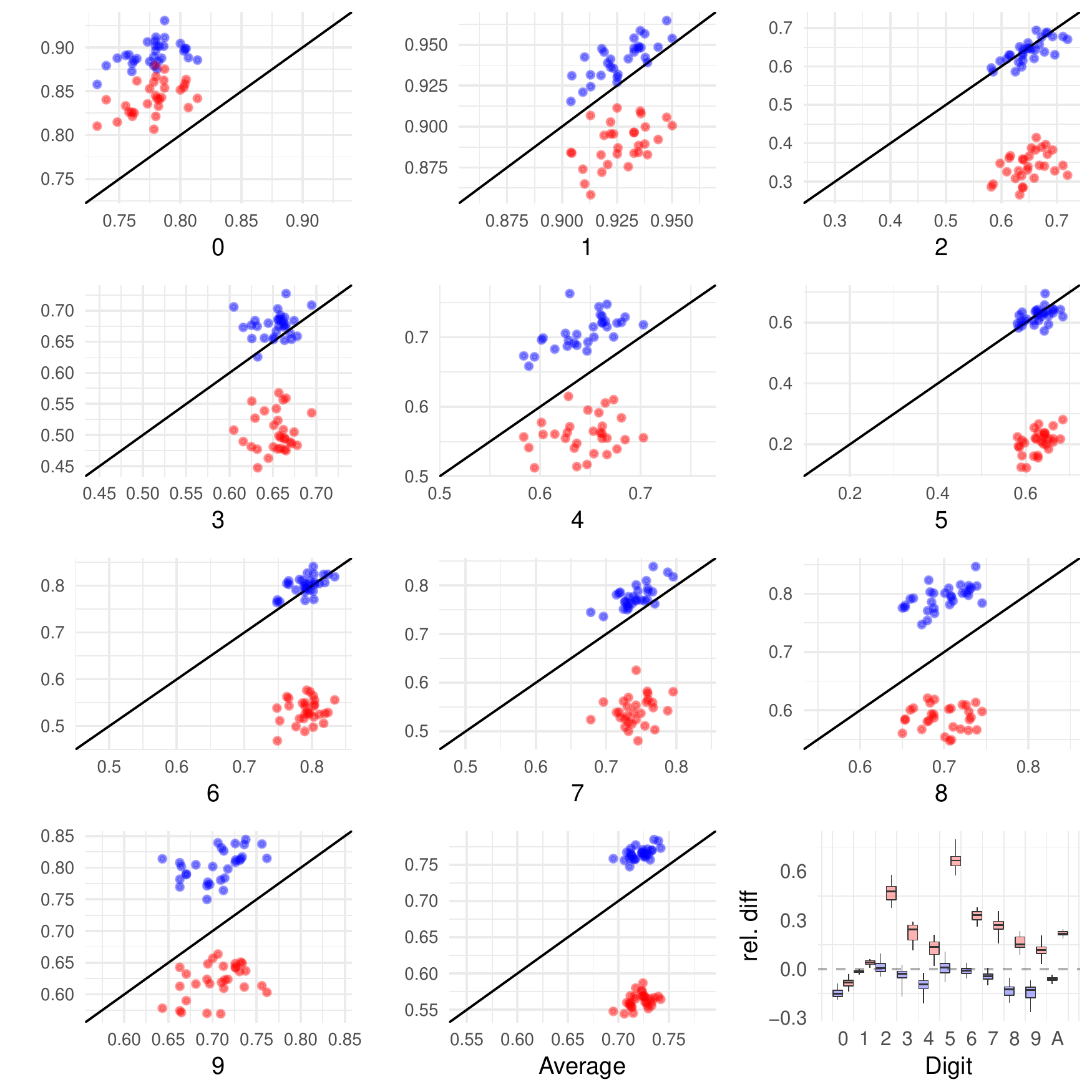} 
\caption{Each plot corresponds to the average $F_1$ class score across a 30-fold cross validation on the full MNIST training data. The horizontal axis represents $F_1$ scores for a random forest classifier trained using dimension 0 features. The vertical axis represents $F_1$ scores from the same random forest trained on different feature sets, one using only dimension 1 features (red points), and the other using dimension 0 and dimension 1 features (blue points). The boxplots show the distribution of the relative differences on $F_1$ scores between training sets.}
\label{plots}       
\end{figure}

Figure \ref{plots} plots the class and overall $F_1$ scores of the random forest over a 30-fold stratified cross validation over the full MNIST training set. In these plots, each blue point is paired with a red point as they share the same first coordinate (i.e. $F_1$ score from training on dimension 0 features). Several important observations can be made based on these plots. The first is that dimension 1 features certainly improve the classification performance of the random forest as evidenced by the significant majority of blue points plotted above the diagonal in many of the classes, and hence in the overall classification. This is most impressed in the class scores for the digits 0, 4, 8, and 9. We verify this observation and measure the improvement introduced by performing a paired t-test using the cross validation scores. We report the result of this test in Table \ref{ttest1}.

\begin{table}[h]
\centering
\caption{Paired t-test of significant difference of $F_1$ scores between training on dimension 0 features and when supplemented with dimension 1 features.}
\label{ttest1}       
\begin{tabular}{cccccccccccc}
\hline\noalign{\smallskip}
Digit & 0 & 1 & 2 & 3 & 4 & 5 & 6 & 7 & 8 & 9 & Overall   \\
\noalign{\smallskip}\hline\noalign{\smallskip}
Mean Diff. & 0.115 & 0.014 & -0.009 & 0.023 & 0.063 & -0.007  & 0.010 & 0.036 & 0.094 & 0.098 & 0.044 \\
$p$-value & 2e-16 & 1e-10 & 0.05 & 3e-05  & 5e-15 & 0.2 & 0.01 &3e-11 & 2e-16& 2e-16&2e-16 \\
\noalign{\smallskip}\hline
\end{tabular}
\end{table}
 
Now, the fact that in all except one class, red points in Figure \ref{plots} are plotted below the diagonal, shows that the random forest performed better when trained using only dimension 0 features than when trained using dimension 1 features. This is confirmed by the red boxplots in the last subfigure showing the distribution of positive relative differences in $F_1$ scores. In addition, the significant vertical distance of the blue points from the red points illuminates the significant increase in classification performance across all classes when dimension 0 features are supplemented to dimension 1 features. We again verify this observation and measure the improvement introduced by performing a paired t-test using the cross validation scores and report the result of this test in Table \ref{ttest2}. All these suggest that dimension 0 features are able to pick up shape characteristics within the digits that are better discriminants for the digit classes and/or are not seen by dimension 1 features.
\begin{table}[h]
\centering
\caption{Paired t-test of significant difference of $F_1$ scores between training on dimension 1 features and when supplemented with dimension 0 features.}
\label{ttest2}       
\begin{tabular}{cccccccccccc}
\hline\noalign{\smallskip}
Digit & 0 & 1 & 2 & 3 & 4 & 5 & 6 & 7 & 8 & 9 & Overall   \\
\noalign{\smallskip}\hline\noalign{\smallskip}
Mean Diff. & 0.050 & 0.051 & 0.299 & 0.172 & 0.148 & 0.415  & 0.268 & 0.235 & 0.206 & 0.182 & 0.203 \\
$p$-value & 2e-16 & 2e-16 & 2e-16 & 2e-16  & 2e-16 & 2e-16 & 2e-16 &2e-16 & 2e-16& 2e-16&2e-16 \\
\noalign{\smallskip}\hline
\end{tabular}
\end{table}

Finally, we highlight that while dimension 0 and 1 features both benefit when one is supplemented by the other, the increase in performance of the random forest is generally much more significant when dimension 1 features are supplemented by dimension 0 features.
 
\subsection{Predicting Higher Dimensional Shape Characteristics}
\label{predict}
We now attempt to explicitly recover higher dimensional shape characteristics using only clustering data. We perform this task in two ways. In the first approach, we use the same random forest classifier in the previous section trained only on dimension 0 features and replace the labels with the number of topological holes to see if the model can learn to classify the digits according to this higher dimensional characteristic. The classes are defined as follows: the digits 1, 2, 3, 5, and 7 have 0 topological hole, the digits 0, 4, 6, and 9 have 1 topological hole, and the digit 8 alone has 2 topological holes. In this approach, the assigned class provides the predicted value for the higher dimensional characteristic. In a second approach, we train a random forest regressor with 1000 trees on the same dimension 0 features and use each of the 50 dimension 1 features in the previous section as labels in an attempt to predict their values. In both approaches, we perform a stratified 30-fold cross validation on the full MNIST training data. 

Table \ref{predclass} shows that the random forest classifier is able to classify at a respectable level of accuracy the digits 0-9 according to the number of topological holes present in their shape using only features based on observed clustering behavior. We remark that this is somewhat expected considering the relatively good performance of the random forest trained on the same feature set in classifying all 10 digits themselves as seen in the previous section. Indeed, the increase in overall $F_1$ score could be logically attributed to the pooling together of digits with the same topological shape as it essentially filters out some confusion introduced when digits that are topologically equal, e.g. the digits 6 and 9, are labelled differently. Nevertheless, this does verify that clustering behavior is indeed able to capture artefacts of the identified higher dimensional characteristic. 

\begin{table}[h]
\centering
\caption{$F_1$ scores over a 30-fold stratified cross validation of the Random Forest classifier trained on dimension 0 features to classify based on the number of topological holes.}
\label{predclass}       
\begin{tabular}{ccccc}
\hline\noalign{\smallskip}
\# of Holes: & 0 & 1 & 2 & Overall \\
\noalign{\smallskip}\hline\noalign{\smallskip}
Mean & 0.842 & 0.816 & 0.634 & 0.764 \\
SD & 0.008 & 0.009 & 0.031 & 0.014 \\
\noalign{\smallskip}\hline
\end{tabular}
\end{table}

Now, to flesh out in more detail the kind of higher dimensional characteristics that intrinsic hierarchical clustering behavior is able to recover, and the level of precision it is able to do so, we examine the performance of the random forest regressor in predicting each of the persistence-based dimension 1 features. For this part, we do a pre-training of the regressor to identify which 50 out of 150 dimension 0 features are best suited in predicting each dimension 1 feature. Hence our training sets in this exercise are dimension 1 feature-specific.
 
\begin{figure}[h]
\centering
  \includegraphics[height=0.3\textwidth,width=0.8\textwidth]{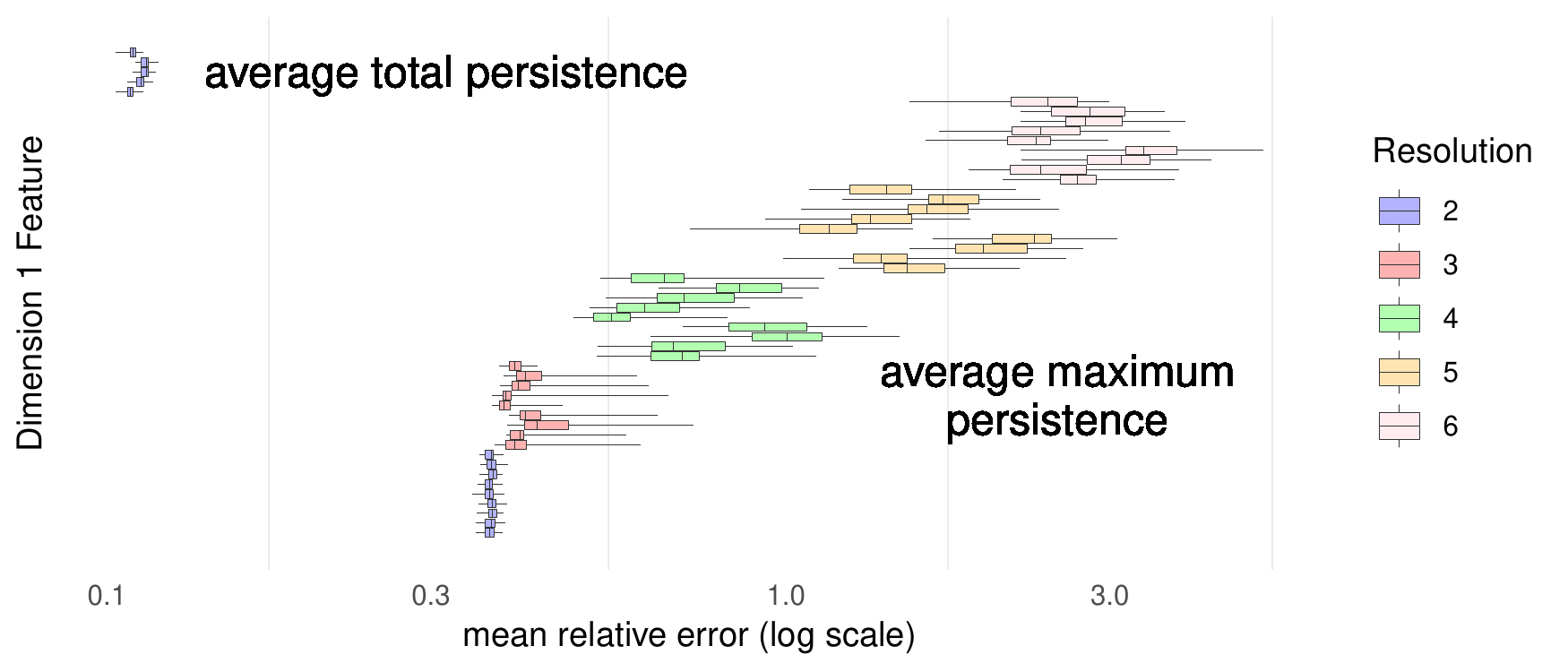} 
\caption{Distributions of mean relative errors from the prediction of the top 50 dimension 1 features by a random forest regressor trained on dimension 0 features. Each boxplot is obtained using scores from a 30-fold cross validation.}
\label{predreg}       
\end{figure}

Figure \ref{predreg} shows the distribution of mean relative errors in the predicted values for each dimension 1 feature across a 30-fold cross-validation. The predicted features are stacked from bottom to top based on their original ranking discussed in the previous section. It is clear that the regressor performs best when predicting features induced from sampling at best resolution, and that the performance and prediction variability decay with resolution. It can also be observed that the distributions of mean relative error for samples referenced from the same landmark point follow a consistent trend at resolutions 3 and above. 

\begin{figure}[h]
\centering
  \includegraphics[height = 0.4\textwidth,width=0.9\textwidth]{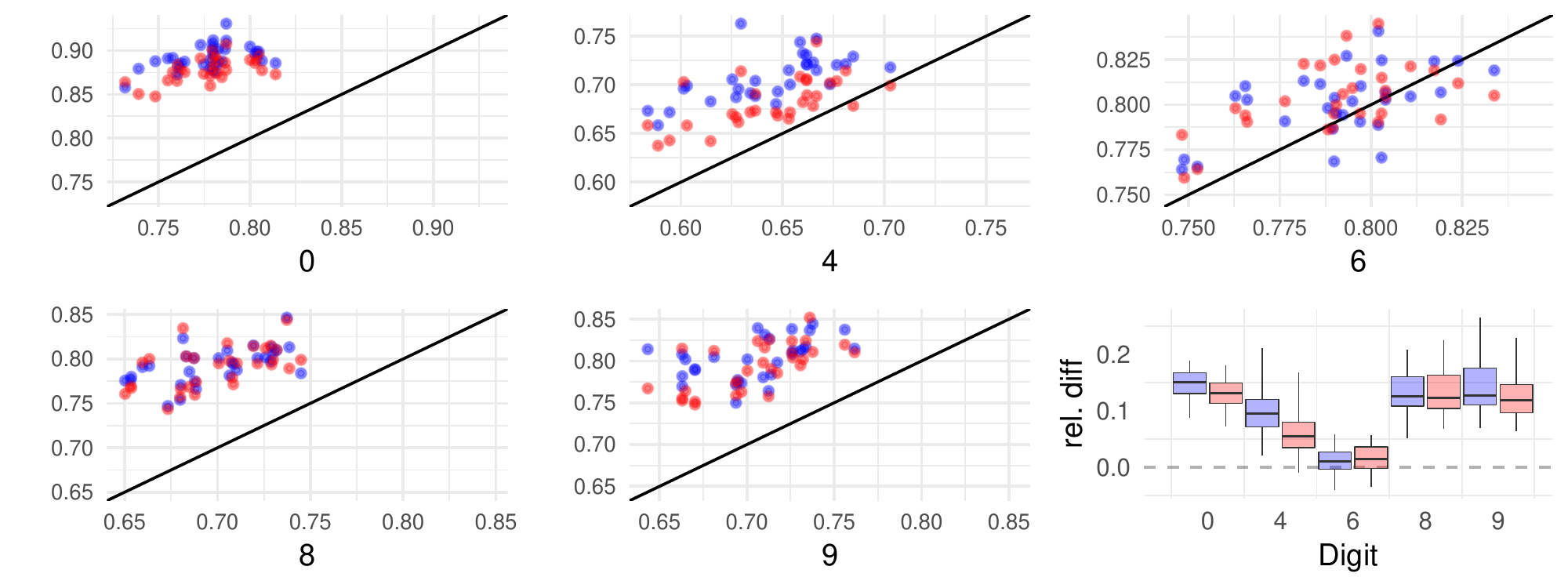} 
\caption{The blue points are the same points from Figure \ref{plots} representing the comparison between $F_1$ scores obtained when a random forest classifier is trained on dimension 0 features (horizontal component) and when the same training set is supplemented with dimension 1 features (vertical component). The red points have the same horizontal component, but the vertical component represents the $F_1$ scores when the dimension 0 features are supplemented with predicted dimension 1 features. The boxplots show the distribution of the relative differences in $F_1$ scores between training sets.}
\label{predreg2}       
\end{figure}

It is interesting to note, however, that despite this apparent significant deviation of predicted values from actual ones, the regressor is still able to use these predicted values to recover a significant chunk of the performance boost reported in Table \ref{ttest1} where actual dimension 1 features are introduced. In particular, this is most notable for the digits 0, 4, 6, 8, and 9 as observed from Table \ref{ttest3} and Figure \ref{predreg2}, further supporting our hypothesis that our dimension zero features, crafted from an observation of clustering behavior, do detect artefacts of, and can be tapped by machine learning algorithms to recover higher dimensional characteristics.

\begin{table}[h]
\centering
\caption{Paired t-test of significant difference of $F_1$ scores obtained when training on dimension 0 features from when the same training set is supplemented with predicted dimension 1 features.}
\label{ttest3}       
\begin{tabular}{cccccccccccc}
\hline
Digit & 0 & 1 & 2 & 3 & 4 & 5 & 6 & 7 & 8 & 9 & Overall   \\
\hline
Mean Diff. & 0.101 & -0.009 & -0.017 & 0.004 & 0.038 & -0.021  & 0.012 & -0.009 & 0.092 & 0.088 & 0.028 \\
$p$-value & 2e-16 & 2e-4 & 0.001 & 0.5  & 2e-9 & 7e-4 & 0
.002&0.02 & 2e-16& 2e-16&2e-16 \\
\hline
\end{tabular}
\end{table}

\section{Conclusion}

In this paper, we showed that specific artefacts of higher dimensional shape information can be recovered from intrinsic clustering behavior observed from samplings of an underlying point cloud. By employing a multi-referenced and multi-resolution sampling of the point cloud, we are able to capture intrinsic clustering behavior from multiple perspectives and levels of granularity. In addition, our application of several hierarchical clustering algorithms provided a nuanced observation of clustering behavior from these sampling settings. Tapping the bottleneck distance provided a holistic way to compare and measure differences between observed clustering behavior. Finally, we showed that statistical features derived from these computed differences can be used to train a machine learning algorithm to classify point cloud objects with higher dimensional shape characteristics, and to produce approximations of these quantities that can be used as proxies for classification tasks. 



\begin{thebibliography}{99}


\bibitem{perea}
	Perea, J.A., Harer, J., \emph{Sliding Windows and Persistence: An Application of Topological Methods to Signal Analysis}, Found Comput Math, 15, pp. 799--838, 2015

\bibitem{ignacio}
 	Ignacio, P.S., Dunstan, C., Escobar, E., Trujillo, L., Uminsky, D., \emph{Classification of Single-Lead Electrocardiograms: {TDA} Informed Machine Learning}, 2019 18th IEEE International Conference On Machine Learning And Applications (ICMLA), pp. 1241-1246, 2019
	
\bibitem{bubenik}
	Bubenik, P., Hull, M., Patel, D., Whittle, B., \emph{Persistent homology detects curvature}, Inverse Problems, 36(2), 2020

\bibitem{lecun}
LeCun, Y., Cortes, C., \emph{The MNIST dataset of handwritten digits}, 1999
	

\bibitem{garin}
  Garin, A., Tauzin, G.,\emph{A Topological ``Reading'' Lesson: Classification of MNIST using TDA}, 2019 18th IEEE International Conference On Machine Learning And Applications (ICMLA), 2019
  
  
\bibitem{jain}
	Jain, A.K., Dubes, R.C., \emph{Algorithms for Clustering Data}, Prentiss-Hall Advanced Reference Series, Prentiss-Hall, Inc., New Jersey, 1998

 \bibitem{leyda}
Almod\'ovar, A., \emph{Studying brain networks via topological data analysis and hierarchical clustering}, PhD Thesis, The University of Iowa, 2016

\bibitem{lumawig}
Ignacio, P.S., Bulauan, J., Uminsky, D., \emph{\textsc{Lum\'awig}: An Efficient Algorithm for Dimension Zero Bottleneck Distance Computation in Topological Data Analysis}, arXiv:2010.00371

\bibitem{roadmap}
    Otter, N., Porter, M.A., Tillmann, U., et al., \emph{A roadmap for the computation of persistent homology}, EPJ Data Sci., 6(17), 2017
	
\end{thebibliography}
\end{document}